# Femtosecond Laser Processing of Germanium: An Ab Initio Molecular Dynamics Study

Pengfei Ji and Yuwen Zhang[1]

Department of Mechanical and Aerospace Engineering

University of Missouri

Columbia, MO 65211, USA

## Abstract

An ab initio molecular dynamics study of femtosecond laser processing of germanium is presented in this paper. The method based on the finite temperature density functional theory is adopted to probe the structural change, thermal motion of the atoms, dynamic property of the velocity autocorrelation, and the vibrational density of states. Starting from a cubic system at room temperature ($300\ K$) containing 64 germanium atoms with an ordered arrangement of $1.132\ nm$ in each dimension, the femtosecond laser processing is simulated by imposing the Nose Hoover thermostat to the electronic subsystem lasting for $\sim 100\ fs$ and continuing with microcanonical ensemble simulation of $\sim 200\ fs$. The simulation results show solid, liquid and gas phases of germanium under adjusted intensities of the femtosecond laser irradiation. We find the irradiated germanium distinguishes from the usual germanium crystal by analyzing their melting and dynamic properties.

---

[1] Corresponding author. Email: zhangyu@missouri.edu

## Nomenclature

| | |
|---|---|
| c | conversion coefficient, $eV/K$ |
| d | length of the simulation system in one direction, Å |
| $k_B$ | Boltzmann constant, $1.38 \times 10^{-23}\ J/K$ |
| E | atomic-scale energy, Ha |
| $E_{II}$ | ion-ion Coulomb energy, Ha |
| $f_i$ | Fermi-Dirac occupation number |
| J | femtosecond laser fluence, $J/cm^2$ |
| M | atomic mass, a. u |
| $N_e$ | number of electrons |
| $N_I$ | number of ions |
| $\boldsymbol{r}_i$ | position of the atom $i$ |
| $\boldsymbol{R}$ | ionic position vector, Bohr |
| T | temperature, $K$ |
| v | velocity, $Bohr/a.t.u$ |
| V | potential term in the Hamiltonian, Ha |
| $\boldsymbol{V}_i$ | three components of the velocity $v_x(t), v_y(t)$ and $v_z(t)$ |

Greek

| | |
|---|---|
| $\beta$ | electron temperature parameter |
| $\mu$ | chemical potential, Ha |
| $\phi$ | Hartree potential, Ha |
| $\psi_i$ | one-electron eigenstates |
| $\mathcal{F}$ | electronic free energy, Ha |
| $\Omega$ | ground potential for an interacting spin Fermi gas, Ha |
| $\mathcal{H}$ | one electron Hamiltonian |
| $\nu$ | vibrational frequency of atom, $s^{-1}$ |
| $\rho$ | electron density |

*Subscripts and Superscripts*

| | |
|---|---|
| $xc$ | exchange and correlation |
| $e$ | electron |
| $I$ | ion |
| $vol$ | volume of the modeled system |

# 1. Introduction

Femtosecond laser processing has been widely used in the fabrication process of cutting-edge semiconductor integrated circuits (ICs) and nanoelectromechanical systems (NEMS). The photolithography, serving as a conventional approach to fabricate the microstructures, has been a predominant technique since the emergent of semiconductors. However, the traditional photolithography suffers from the drawbacks of high cost and complex treatment for the masks with the miniaturization of structure due to optical diffraction in nanoscale. The femtosecond laser processing (nanolithography), due to its inherent advantages of maskless super resolution [1], offers a potential solution for the high mask cost. In addition, compared with the nanosecond pulses, the femtosecond laser has the advantages of high peak power intensity and a relatively smaller heat-affected zone, which make it ideal for a wide range of applications in other fields of material sciences [2–6]. In many engineering applications ranging from laser micromachining to surface treatment, the femtosecond laser material interaction has become an increasingly hot topic. The laser intensity of femtosecond laser can even be up to $10^{21}\ W/m^2$ [7], which results in the breaking down of traditional phenomenological laws. Hence, for the purpose of achieving a better comprehension and understanding of the femtosecond laser nanolithography, it is of great necessity to investigate the melting and dynamic effects relating with energy transfer and conversion in the atomic scale.

The laser material interaction has attracted considerable attentions in the past two decades due to both the relevant technological importance and theoretical interest. Because the characteristic times of the energy carriers are comparable to the characteristic energy excitation time, there are two competing interpretations to the thermal mechanical phenomena: one is the plasma annealing, another one is thermal annealing. The plasma annealing describes the microscopic mechanism as the lattices of semiconductors directly transform into a disordered state long before the system become thermally excited. For semiconductors, like silicon, phonons are the major energy carriers. Because the free electron density in semiconductor is much lower than that in metal, the participation of electrons can be neglected. The mechanism under femtosecond laser irradiation is generally regarded as an athermal transition, which mainly softens the interatomic bonds in the time domain that is long before the conventional thermal transition caused by heat transfer from the electrons to phonons [8]. After femtosecond laser irradiation, the electrons transform from the bonding to antibonding states by depleting the bond charges. In other words, when the stage of disordering of the latter happens, the ion temperature still remains at room temperature. A tight-binding model was introduced to study the effects of a dense electron-hole plasma on the instability of carbon, silicon and germanium [9,10]. The other alternative explanation is the thermal melting, which requires the direct energy transfer between the excited electrons and the ions. Due to the electron-phonon coupling, the large amount of excited electron energy transfers to ions within the picoseconds time domain, which causes thermal melting of the lattice.

For the femtosecond laser, the pulse duration is shorter than the electron-lattice thermalization time (in the order of picosecond). Because the excited electrons do not have sufficient time to transfer a large amount of energy to the ions, plasma annealing seems to be the dominating mechanism. The ultrashort laser pulse makes it possible to excite the electronic states of a solid long before the appreciable energy transporting to the lattice vibrational states. Shank *et al.* [11] showed transition from crystalline order to disorder with melting of the surface occurred in less than 1 $ps$ under 90 $fs$ laser illumination on silicon. Rousse *et al.* [12] reported the nonthermal melting process in InSb at femtosecond resolution by using a ultrafast time resolved X-ray diffraction. An atomic-scale visualization of internal dynamics presented by Lindenberg *et al.* [13] demonstrated the nonthermal transition from crystalline solid to disordered liquid. Apart from experimental investigation, by employing first principles calculation, Zijlstra *et al.* [14] showed the nonthermal melting of InSb induced by femtosecond laser based on dramatic changes of transverse acoustic phonons at the boundary of Brillouin zone. Subsequently, anharmonic effects which dominate the atomic motions for times greater than $100\ fs$ were studied [15]. Moreover, the relationship of thermal phonon squeezing under low fluence femtosecond laser irradiation and the same pathways of atomic motion at the first stages of nonthermal melting was reported by Zijlstra e*t al.* [16], which originated from the femtosecond laser induced softening of the interatomic force constant. The concluded that the phonon squeezing playing the role as the precursor to nonthermal melting with the increasing of laser fluence. Threshold intensities of nonthermal phase transition for semiconductors (such as, Si and InSb) were reported by utilizing the semiclassical electron-radiation-ion dynamics (SERID) [17]. In this work, the ab initio molecular dynamics simulation method based on finite temperature (FT-DFT) is used to simulate femtosecond laser interaction of germanium. The approach has already been employed to simulate the laser melting of silicon and graphite [18,19]. Since the FT-DFT method incorporates self-consistently the effect of thermalized electronic excitations and fractionally occupied state, it can reasonably and accurately describe the electronically hot system. The original work of ab initio molecular dynamics of excited electrons was presented by Alavi *et al.* [20]. We carried out simulation with different incident laser intensities of $\sim 100\ fs$ time duration. All the cases started from germanium crystal at room temperature $(300\ K)$. To obtain the final equilibrium state of the electronic and ionic subsystem, we monitored the temperature of nucleus as output during the entire simulation process. The phase changes from the solid state to liquid state (melting point of Ge: $1211.40\ K$) and gas phase (evaporation point of Ge: $3106\ K$) were validated from the microstructure property, thermal motion and dynamics feature analysis.

## 2. Theory and Method

### 2.1 Ab Initio Molecular Dynamics Based on FT-DFT

The concept of finite temperature version of the density functional theory was originally proposed by Mermin [21] when it referred to the thermal properties of electron gas. The functional introducing the finite electron temperature of effects into the DFT is

$$\Omega[\boldsymbol{\varphi}_i(\boldsymbol{r})] = E^{KS}[\boldsymbol{\varphi}_i(\boldsymbol{r})] - T_e S \tag{1}$$

where $E^{KS}[\boldsymbol{\varphi}_i(\boldsymbol{r})]$ is the Kohn-Sham energy functional at $T_e = 0$ and $S$ is the entropy, which can be expressed as

$$S = -k_B \sum_i [f_i \ln f_i + (1 - f_i) \ln(1 - f_i)] \tag{2}$$

For the specific case of non-interacting Fermions, the grand potential expression is

$$\Omega(\mu VT) = -k_B T det^2 (1 + e^{-\beta(\mathcal{H}-\mu)}) \tag{3}$$

where $\mu$ refers to the chemical potential acting on the electrons and $\mathcal{H}$ is the one electron Hamiltonian $\mathcal{H} = -\frac{\hbar^2}{2m}\nabla^2 + V(\boldsymbol{r})$. The power of two is because of the spin multiplicity. The effective density dependent potential term in Hamiltonian is expressed as

$$V(\boldsymbol{r}) = -\sum_I \frac{Z_I e^2}{|\boldsymbol{R}_I - \boldsymbol{r}|} + V_H + \frac{\delta \Omega_{xc}}{\delta \rho(\boldsymbol{r})} \tag{4}$$

The free energy $\mathcal{F}$ is a functional of the electron density $\rho(\boldsymbol{r})$, the Fermi-Dirac occupation number $f_i$ of the one-electron eigenstates $\boldsymbol{\varphi}_i(\boldsymbol{r})$ and the chemical potential $\mu$. Recalling the mathematical relationship $\left(\frac{\partial \Omega}{\partial \mu}\right)_{\rho(\boldsymbol{r})} = -N$ in thermodynamics, we can get the average electron number. Therefore, the sum of the Helmholtz energy [20] of one electron of density $\rho(\boldsymbol{r})$ and the ion-ion Coulomb energy $E_{ions}(\boldsymbol{R}^N)$ is

$$\mathcal{F} = \Omega + \mu N_e + E_{ions}(\boldsymbol{R}^N) \tag{5}$$

where $\Omega$ denotes the grand potential for an interacting spin Fermi gas within DFT. In order to obtain the correct total electronic free energy of the interacting electrons, the extra terms have to be included in $\Omega^{KS}$, namely

$$\Omega = -\frac{2}{\beta}\ln\det\left(1 + e^{-\beta(\mathcal{H}-\mu)}\right) - V_H - \int \rho(\boldsymbol{r})\left(\frac{\delta\Omega_{xc}}{\delta\rho(\boldsymbol{r})}\right)d\boldsymbol{r} + \Omega_{xc} \tag{6}$$

where $\beta = \frac{1}{k_B T_e}$ refers to as the electron temperature parameter and $\Omega_{xc}$ is the finite temperature exchange correlation grand potential functional.

The electron density is

$$\rho(\boldsymbol{r}) = \sum_i^{occ} f_i(T)|\boldsymbol{\varphi}_i(\boldsymbol{r})|^2 \tag{7}$$

and the occupation number is

$$f_i(T) = \frac{f_i(0)}{1 + e^{-\frac{1}{k_B T_e}(\epsilon_i^{KS}-\mu)}} \tag{8}$$

where $\epsilon_i^{KS}$ is the Kohn-Sham eigenvalues and $f_i(T)$ is the Fermi-Dirac occupation number in the grand canonical ensemble.

The optimization for the specific atomic configuration can be realized by adopting the Lanczos algorithm [22]. After the electron density $\rho(\boldsymbol{r})$ is obtained, the dynamic simulation is performed based on the optimization of the electron density at every time step. Then, the Hellmann-Feynman forces acting on the nuclear can be calculated via the differentiation of $\mathcal{F}$ with respect to the ionic coordinates.

$$\boldsymbol{F}_I = -\nabla_I \mathcal{F} \tag{9}$$

Eventually, we can perform molecular dynamics simulation for the system with various temperatures of electrons.

## 2.2 Modeling and Simulation

The ab initio molecular dynamics simulation was started with germanium crystal at room temperature ($300\ K$). The CPMD package 3.15.3 [23] based on plane wave peseudopotential implementation of FT-DFT and incorporating self-consistently the effects of thermal electronic excitations and fractionally occupied states was used. As introduced in Section 2.1, certain method is particularly congruent to deal with the electronically hot problems. A cubic system containing 64 germanium atoms were modeled within $11.316\ Å$ at $x$-, $y$- and $z$- directions, which corresponds to the density of germanium at $5.323\ g/cm^{-3}$. The Γ point was used to sample the Brillouin zone of the molecular dynamics supercell. The simulation time step was set $4\ a.u.$ ($\sim 0.096,755\ fs$) to obtain integration of the ionic degrees of freedom. The exchange correlation in our simulation was represented by the local density approximation (LDA) [24]. As pointed out by Silvestrelli et al. [25], the LDA underestimates the energy gap and the Γ point of the supercell used in FT-DFT simulation merely gives a poor representation of the density state condition band. Nevertheless, the LDA well describes the antibonding character of the conduction states and only grass features of the conduction band are relevant at high levels of excitation. We adopted the norm conserving pseudopotentials with Stumpf-Gonze-Schaeffler [26] pseudolization method [27]. Periodic boundary conditions were applied in all three directions of the primitive cell of diamond cubic crystal structure. In our simulation, all cases adopted the cutoff of $35\ Ry$ to expand the electronic orbitals in plane waves. The single particle electronic states were set as 200 to take all the levels of electronic excitation into consideration.

In order to perform FT-DFT simulation, the wavefunction optimization and geometric optimization were performed first to calculate the electronic structure of the system and the optimization results showed the cutoff of $35\ Ry$ is sufficient engough to get the converged energies. After obtaining the optimized electronic structure, all the cases were run for 1,000 time steps at room temperature ($300\ K$) with Nose Hoover [28] thermostats imposed on each degree of freedom for both electrons and nuclei, so that the electronic subsystem and ionic subsystem could reach to sufficient equilibrium state before laser irradiation.

The simulation time of femtosecond laser irradiation locates in the subpicosecond domain, which corresponds to a nonequilibrium population of the excited "hot" electrons with a still "cold" atomic lattice [29]. In our simulation, due to the fact that the estimated relaxation time for the electron-electron interaction ($\sim 10\ fs$) is much shorter than the electron-ion interaction ($\sim 1\ ps$) and ion-ion interaction ($\sim 10\ ps$), the electrons subsystem was assumed to achieve high level temperature instantaneously, during the period was less than $1\ ps$ since laser irradiation. However, the ionic subsystem was treated tardily respond to the irradiation and let it run freely to reach final equilibrium with the electronic subsystem. We can find the same kinds of approximation and simulation approach simulating the scenario of femtosecond laser irradiation, which uses the constant electron temperature control to characterize the instantly excited electronic subsystem. Two typical treatments adopting the aforementioned methodology are laser melting of graphite [18] for $0.5\ fs$ and silicon [19] for $0.4\ fs$. Besides using the free energy functional approach, simulation technique of elevating electron temperature as a constant value to represent the excited electronic subsystem, can also be seen in the investigation of crystal stability in the density functional perturbation theory calculations [8], potential energy surface energy calculation of InSb in dealing with ultrafast nonthermal melting [15] and ab initio molecular dynamics simulation of femtosecond laser interaction with vitreous silica [30]. In addition, as reported in the estimations of experiments, the electron temperature can be as high as $100,000\ K$ [18]. Since femtosecond laser creating electron-hole pairs by electron excitation on a time scale shorter than the time of ionic motion [31], we carried out simulation of femtosecond laser irradiation by instantly increasing the electron temperature to $20,000\ K$, $25,000 K$, $30,000 K$, $35,000 K$, $40,000 K$, $50,000 K$ and $60,000 K$ from the $1,001 th$ to $2,000 th$ time steps, which corresponded to the time duration of $\sim 100\ fs$. A formal justification of the electron temperature $T_e$ to characterize the femtosecond laser excitation can be found in [21] and the forces acting on the ions are the first derivative of the electronic free energy [20]. In order to observe the further results of femtosecond laser processing of germanium, we carried on the simulation with constant total energy lasting for another 2,000 time steps by removing the energy control of electronic subsystem.

The evaluation of electron temperature was based on the Maxwell-Boltzmann distribution of electron gas

$$\langle E \rangle = \frac{3k_B}{2}\langle T \rangle = c\langle T \rangle \tag{10}$$

where $c$ is a conversion coefficient from electron temperature of $1\ K$ to average electron energy $8.6173 \times 10^{-5} eV$. From the engineering prospective, when it refers femtosecond laser material processing, laser pulse duration and laser fluence (intensity) are two crucial parameters to characterize femtosecond laser. In this study, the instant incident of laser fluence can be expressed by the energy increment of electron subsystem:

$$J = \frac{N_e N_I c \Delta T}{d_x d_y} \qquad (11)$$

where $N_I$ and $N_e$ refer to the number of atoms included in the system and the number of electrons circling around one germanium nuclei; $c$ is the conversion coefficient from electron temperature to average electron energy; $\Delta T$ is the electron temperature increase induced by laser irradiation; and $d_x$, $d_y$ are the width and length of the modeled system. The calculated average electron energies, laser fluences and laser intensities are listed in Table 1.

## 3. Results and Discussion

### 3.1 Ion Temperature Evolutions

We calculated the temperatures of the ionic subsystems along with the MD simulation for all the cases. As is shown in Fig. 1, the ion temperatures exceed the melting point and evaporation point under the electron irradiation conditions at temperatures of $25,000\ K$ and $50,000\ K$. We calculated the temperature of ions according the following definition

$$T_I = \frac{2\bar{E}_I}{3(N-1)k_B} \qquad (12)$$

As we can see from Fig. 1, the lattices remain at relatively lower temperatures in the first $1,000$ time steps of laser irradiation. When the condition of femtosecond laser interaction is imposed, the relatively low temperature states of ions keep about $200$ time steps ($\sim 0.2\ fs$) and then are followed by slow temperature increments from 1,200 to 1,500 time steps. After 1,500 time steps, the ion temperatures show abrupt increases, especially for cases of the electronic subsystems with relatively higher energies. The advents of ion temperatures exceeding the melting and boiling points are also early for the electron systems with higher temperatures than those with lower electron temperatures. After the temperatures of ion subsystems increasing to certain values, they stayed and oscillated in small ranges at temporarily equilibrium states with the electron subsystems. The same sort of kinetic energy (ion temperatures) responses to various intensities of femtosecond lasers are also reported in [29]. The temperature curve ($1,000 - 4,000$ time steps) for the case of $20,000\ K$ is distinctly different from that of $25,000\ K$, which indicates that for the case of $20,000\ K$ femtosecond laser almost cannot make the ions to break most of the valence bonds connecting them but only keep with wild oscillations around their equilibrium positions. The same condition of femtosecond laser distinct threshold of $0.15\ J/cm^2$ that enables the strong excitation and nonthermal process was experimentally measured for femtosecond processing of gallium arsenide [32]. It is not definite to draw the final conclusion that the ions will still keep at low temperatures for the case of electron temperature $20,000\ K$, with the time duration to picosecond time scale. As pointed in Lindemann's law [33], once the vibrational amplitude reaches up to 10% of the nearest neighbor distance, the vibration will disrupt the equilibrium crystal lattice and the melting phenomenon eventually takes place. The short simulation period could not reveal the melting phenomena for this case. Nevertheless, we can find that even though the simulation process lasts for $\sim 0.3\ ps$, the temperature evolutions for the seven cases still distinguish from each other and the ion temperatures are much lower than the corresponding electron temperatures for all the cases. The phenomena can be interpreted from two aspects. On one hand, the incident energy heats the electrons more strongly than the ions. On the other hand, atoms and ions are much heavier than electrons, which lead to the inefficient thermal energy transfer in a two-body collision because the masses are dissimilar. Moreover, as computed in Table 1, the laser fluence exciting electron temperature to $60,000\ K$ is $0.1316\ J/cm^2$, which is in the same order of magnitude as reported for the femtosecond laser ablation threshold fluence ($0.2\ J/cm^2$) for silicon [34] and $0.17\ J/cm^2$ for graphite films [35].

**3.2 Free Energy Calculation**

The quantities of free energies of the simulation systems were recorded to get a further insight of the changes of the microsystem. The Fermi-Dirac distribution was used for the purpose of filling the electronic states as a function of the temperature in the free energy functional calculation. Figure 2 shows the free energies of the system under different femtosecond laser irradiations. With the increasing laser energy intensities, the free energy drops greater ranges and stays at lower energy values. To get the results more explicitly, the detailed evolution curves of each case are shown in Figs. 3. The curves for free energies correspond to the ion temperature profiles shown in Fig. 1. It means that while there is a free energy drop (see Fig. 3), there is an increment of the ion temperature (see Fig. 1). This kind of trends can also be found in the simulation result of silicon [25]. The comparisons among Figs. 3(a)-(d) demonstrate that even though the electron temperatures induced by the incident laser are tens to hundreds times higher than the ion temperature, the electronic-ionic energy exchange cannot happen instantaneously and efficiently for some cases that the electron temperatures lower than a criteria (see the $20,000\ K$ and $30,000\ K$ cases). Additionally, the capability of the system to absorb external laser energy is also affected by the structural changes and the dependency of the absorbed energy on the laser intensity is nonlinear [36].

## 3.3 Structural Change due to Electron-Ion Interaction

To understand the thermal effect of femtosecond laser nanolithography, structural change is the primary parameter to evaluate the results of irradiated germanium. The radial distribution function (RDF) $g(r)$ describes how the density varies as a function of distance from a reference particle. It is also referred to as pair distribution function or pair correlation function, which describes how, on average, the atoms in a system are radially packed around each other. As an important structural characteristic of crystal germanium, we computed RDF based on the trajectories of atoms produced at different simulation stages from the initial equilibrium state at room temperature to the subsequent laser irradiation process. In each case, we took the number of $\delta r$ steps as 1,000. From the calculated RDF, the probability to find an atom in a shell $dr$ at the distance $r$ of another atom chosen as a reference point can be obtained. The mathematical expression of the RDF is

$$g(r) = \frac{Ndn(r)}{V_{vol}4\pi r^2 dr} \qquad (13)$$

where $dn(r)$ represents the number of atoms inside a shell region between $r$ and $r + dr$ for a given system. $N$ and $V_{vol}$ are the total number of atoms in the model and its volume, respectively. Another reason that we calculated the RDF is that the internal energy of the system is related to the pair correlation function.

The RDFs for the room temperature were calculated first and the results are shown in Fig. 4. It can be clearly seen that there are peaks and valleys distributing in the range of $0 - 7$ Å, which reflect the regular arrangement of germanium atoms in the crystal at room temperature. In order to obtain the structural information of germanium under irradiations of difference femtosecond laser fluences, we chose the cases with electron temperatures of 20,000 K, 30,000K and 60,000 K to calculate the RDFs for comparison (see Figs. 5-7). The temporarily stable ion temperatures for the three cases are approximately 400K, 1,500 K and 3,500 K, respectively. We also computed the RDFs for germanium in three phases (corresponding to 400K, 1,500 K and 3,500 K) that the electron temperature and ion temperature are the equal (see the purple curves in Figs. 5-7). The maximum distance in real space to sample the number of germanium atoms was chosen to be 7 Å. As pointed out in [29], the electronic configuration does not stay in ground state and ionic positions are no longer in equilibrium, after femtosecond laser irradiation. Therefore, the abrupt shock of high level energy in the electronic subsystem causes the structure of germanium disturbance since the initial irradiation stage of femtosecond laser. Once a sufficiently high proportion of electrons are promoted from bonding to antibonding states, the structure transition will occur. The emergences of broadening peaks and valleys from $1,001-2,000$ time steps to $2,001-3,000$ time step then to $3,001-4,000$ time steps indicate the germanium atoms are gradually ionized, which are also seen in [17] for RDFs of Si and InSb, [29] for GaAs and [35] for the RDFs of graphite film. Because the RDFs in Figs. 5, 6 and 7 represent distinctly differently phase states, the shapes and values correspondingly distinguish from each other. The different degrees of broadened RDFs for the three cases reveal the atomic distances are separated at different stages due to the shock of incident lasers. With the higher laser fluence, the more random distribution of atom occurs (see the valleys in the red curves in Fig. 6 and Fig. 7), corresponding to broader RDF peaks and valleys. But due to the atomic separation and thermal expansion, the maximum peak that shows the greatest probability of finding atoms at short distance consequently moves to larger atomic distance (see the first peak of different stages in Figs. 5, 6 and 7). At the same time stage, the broader RDF peaks and valleys correspond to higher laser fluence.

For the case of electron temperature 20,000 K, comparing the red curve with the green and blue curves, we can see that even though the ion temperature does not rise to the melting point by the hot electrons, the structure has already changed. As seen in Fig. 1, the temporarily stable temperature of the case of 20,000 K is around 400K. Thus, to make the prediction even more persuasive, we calculated the RDF of germanium, whose temperatures of electronic subsystem and the ionic subsystem are both set as 400 K (see the purple curve in Fig. 5). The profile of the purple curve ($T_{e\&I} = 400$ K) in Fig. 5 resembles the curve ($T_{e\&I} = 300$ K) in Fig. 4, which indicates atoms in the solid germanium are kept as ordered arrangements. In addition, the compare the purple curve and the red curve, because the ion temperatures are approximately equal and the irradiation initially starts, the RDF results show the almost overlapping profiles.

## 3.4 Thermal Motion of Atoms

To quantitatively measure the extent of spatial random motion of the atoms, we calculated the mean square displacement (MSD) of the atomic thermal motion for germanium at difference phases with selected stages like Section 3.3. The trajectories caused by random collisions of atoms from each other are very complex during the evolution of a MD simulation process. By adding the square of the distances, the MSD method overcomes the drawback of elimination effects induced by the addition of the positive and negative values counting from a prefixed position. The MSD is mathematically defined by taking the average of the summations of square distance:

$$MSD(t) = \langle r^2(t) \rangle = \langle |r_i(t) - r_i(0)|^2 \rangle \quad \text{Eq. (14)}$$

where $r_i(t)$ represents the position of the atom $i$ at the time $t$ and $\langle \cdots \rangle$ denoted as an average on the time steps and the particles. Figure 8 shows the calculated MSD of the solid state germanium and the value at the end of the 1,000 time steps (96.755 $fs$) is about 0.02 Å$^2$. Because the thermal motion of atoms is only characterized by phonon vibrations, there is no appreciable MSD. In Figs. 9-11, the calculated MSDs of different simulation stages since the femtosecond laser irradiation are plotted. A general tendency of increasing values at the ending point of each 1,000 time steps is revealed in Figs. 9-11. According to [10], the differences of MSDs originated the transition of valence electrons into conduction band and instability of transverse acoustic phonons and longitudinal optical phonons. In addition, for the cases of higher electron temperature, the MSD values are correspondingly higher. The discrepancies of purple curves between the other three curves verify the predication that germanium with equal electron temperature and ion temperature is different from the ones that the electrons are excited. Moreover, the slope of red curve in Fig. 9 is relatively steeper than the purple curve in Fig 9. Thus, we can see that the motion property of the laser irradiated germanium is more like fluids rather than the germanium with same temperature at the solid state. The remarkably MSD under high energy femtosecond laser irradiation is also reported for vitreous silica [30], whose ion temperature is ~500 $K$ under high energy femtosecond laser irradiation of $T_e = 25,000\ K$. In additional, another simulation results of a self-developed code for high excited valence electron system (CHIVES) [37] were reported to have the quantitatively comparable MSDs of silicon under femtosecond laser excitation [16]. The other point that deserves our attention is the red curves in each of Figs. 9-11, when they are compared with the green and blue curves. We can find that in Fig. 9, the final value of red curve MSD is 0.058 Å$^2$. However, the squared distance of thermal motion of the blue one in Fig. 9 (0.130 Å$^2$) is more than 2.241 times of that in red. In addition, the ratio of the final value between blue cure and green curve gradually narrow down from Figs. 9 to 11. In Fig. 11, the blue curve and green curve are almost coincide with each other, which indicates that the nonthermal melting of laser irradiated germanium thoroughly developed in the $2,001 - 3,000$ time steps and there are no significant differences distinguishing from the subsequent $3,001 - 4,000$ time steps. In Fig 9, the red curve of initial electronic excitation is above the purple curves with equal electron and ion temperatures. However, it differentiates from the cases in Figs. 10 and 11, which could be attributed to retard response for lower femtosecond laser fluence.

### 3.5 Dynamic Property of Velocity Autocorrelation Function

The velocity autocorrelation function (VAF) is also an important parameter to characterize the dynamic properties of femtosecond irradiated germanium. It is important in revealing the underlying nature of the dynamic process operating in the simulated molecular system. The VAF is mathematically defined as:

$$Z(t) = \frac{\sum_{i=1}^{N}(\boldsymbol{V}_i(t) \cdot \boldsymbol{V}_i(0))}{\sum_{i=1}^{N}(\boldsymbol{V}_i(0) \cdot \boldsymbol{V}_i(0))} \qquad (15)$$

where $\boldsymbol{V}_i(t)$ includes the three components of the velocity $v_x(t)$, $v_y(t)$ and $v_z(t)$. We computed the VAFs for system with electron temperature at $300\ K$, $20,000\ K$, $30,000\ K$ and $60,000\ K$, repectively. The sampling time period is $1,000$ time steps for all cases. As is seen in Fig. 12, the computed VAFs for the last three cases are always positive, which lead to the high values of the diffusion coefficient. However, the VAF at room temperature contains both the positive part and negative part. The results can be interpreted in terms of the thermophysical interaction of atoms. Firstly, let us consider a single atom at time zero that has a specific velocity $v$. By assuming the atoms in the system do not interact with each other, according to the Newton's Second Law, the atom will retain the initial velocity all the time; the VAF profile will be a horizontal line with the evolution of time. If the forces are small but not negligible, then the magnitude of VAF will change gradually under the influence of the weak forces. Observing the green, blue and purple lines of the $20,000\ K$, $30,000\ K$ and $60,000\ K$ cases, one can see that the VAFs render simple decays, which demonstrates the presence of weak force slowly destroy the velocity correlations. For the solid germanium at room temperature, the atoms are compacted closely to each other. They will search the position at which there is a balance between the attractive and repulsive forces and the energies of the atoms can be regarded stable for such system. The motion of atoms is like dots connected by springs with back and forth oscillations, so the regular position cannot be broken down and disordered. Hence, the calculated VAF show strong oscillations from positive to negative values with periodicity. The red plot in Fig. 12 shows a damped harmonic motion, which verifies that the germanium at room temperature is a usual solid without excited electrons. But for the cases of $20,000\ K$ and $30,000\ K$ germanium after femtosecond laser irradiation, they show the resemble shapes as the $60,000\ K$ germanium. It is contradictory to the definition of phase based on the temperature, but such contradictories prove the validity of the discrepancies of germanium between $T_{e\&I} = 400$ K and $T_I = 400$ K. It is speculated that such $20,000\ K$ and $30,000\ K$ germanium states as meta-states that differentiate from the usual solid and liquid in nature. In addition, we can find the curve of the $60,000\ K$ is the steepest among the three. The most damped curve can be considered there is only one collision between two atoms and then diffuse away, which reflects the chemically covalence bond break due the exited electrons.

### 3.6 Vibrational Density of States

The vibrational density of states (VDOS) can be used to characterize the number of states per interval of energy at each energy level. We calculated the VDOS by performing Fourier transform of the VAF over all atoms. The normalized VDOS of $N$ atoms system is given by

$$g(v) = \int_{-\infty}^{+\infty} \frac{\sum_{i=1}^{N}(V_i(t) \cdot V_i(0))}{\sum_{i=1}^{N}(V_i(0) \cdot V_i(0))} e^{2\pi i v t} dt \qquad \text{Eq. (16)}$$

Thermal properties, such as the heat capacity and thermal conductivity as well as some other material properties are strongly impacted by the vibrational density of states (VDOS). Hence, to get a profound grasp of the laws governing the vibrational properties of femtosecond laser irradiated germanium is of high technological and fundamental interest. Figure 13 indicates there are high diffusive modes in the low frequency regions. But for the case of germanium with $T_{e\&I} = 300$ K at room temperature, the peak evidently distinguishes from the other three. Therefore, we can conclude that the VDOS of the cases have changed due to the phase transitions induced by the incident femtosecond laser.

## 4. Conclusions

An ab initio molecular dynamics study of the thermal and dynamic effect induced by femtosecond laser irradiation is performed in this paper. By employing the finite temperature density functional theory and incorporating with direct energy minimization, the accurate pseudopotential description of core electrons and the Nose Hoover dynamics temperature control of the electronic and ionic subsystem, our simulation successfully obtained the results of the germanium in three phases (solid, liquid and gas). Even though the simulation in the present work is simplified in some extent, we believe that it captures the essential melting and dynamic features induced by different thermal excitation of the electronic subsystems. The unmelt germanium show essentially liquid and gas characteristics that are different from the usual solid germanium without excited electrons around the nucleus. The structure changes and maximum value of the atomic thermal motion depend on the excited levels of the thermalized electrons. The melting and dynamic response of ionic system also depends on the incident laser intensity, which means the higher excited electronics subsystem, the rapider temperature response of the ionic subsystem. However, there are some thresholds for the energy of the electronic subsystem that limits the phase change from solid to liquid and liquid to gas phase. The melting and evaporation are determined by both the electronic and ionic sub systems.

Our work demonstrates the great potential of ab initio MD simulation of the femtosecond laser interaction with semiconductor materials. With the advantages over classical MD simulation that interatomic potential empirically defined and other numerical modeling and simulation methodologies, it is possible to broaden the ab initio MD simulation to thermodynamical problems involving the participation of electron interaction and transport in the future.

## Acknowledgment

Support for this work by the U.S. National Science Foundation under grant number CBET-1066917 is gratefully acknowledged.

**Table Captions**

Table 1 Femtosecond Laser Parameters in the Simulation

**Table 1**

| Computational Case | Electron Temperature ($K$) | Average Electron Energy ($eV$) | Femtosecond Laser Fluence ($J/cm^2$) | Femtosecond Laser Intensity ($10^{12}\ W/cm^2$) |
|---|---|---|---|---|
| 1 | 20000 | 1.7235 | 0.0442 | 0.4566 |
| 2 | 25000 | 2.1543 | 0.0552 | 0.5707 |
| 3 | 30000 | 2.5852 | 0.0663 | 0.6849 |
| 4 | 35000 | 3.0161 | 0.0773 | 0.7991 |
| 5 | 40000 | 3.4469 | 0.0884 | 0.9132 |
| 6 | 50000 | 4.3086 | 0.1104 | 1.1415 |
| 7 | 60000 | 5.1704 | 0.1325 | 1.3698 |

# Figure Captions

Fig. 1   Ion temperatures

Fig. 2   Free energies for all the cases

Fig. 3   Free energy change for different cases

Fig. 4   RDF of germanium crystal at room temperature (300 $K$)

Fig. 5   RDFs at different stages, solid state (20,000 $K$)

Fig. 6   RDFs at different stages, liquid state (30,000 $K$)

Fig. 7   RDFs at different stages, gas phase (60,000 $K$)

Fig. 8   MSD of germanium crystal at room temperature

Fig. 9   MSDs at different stages, solid state (20,000 $K$)

Fig. 10  MSDs at different stages, liquid state (30,000 $K$)

Fig. 11  MSDs at different stages, gas state (60,000 $K$)

Fig. 12  VAFs of cases calculated from 1,000 time steps simulation

Fig. 13  VDOS of cases sample from 1,000 time steps simulation

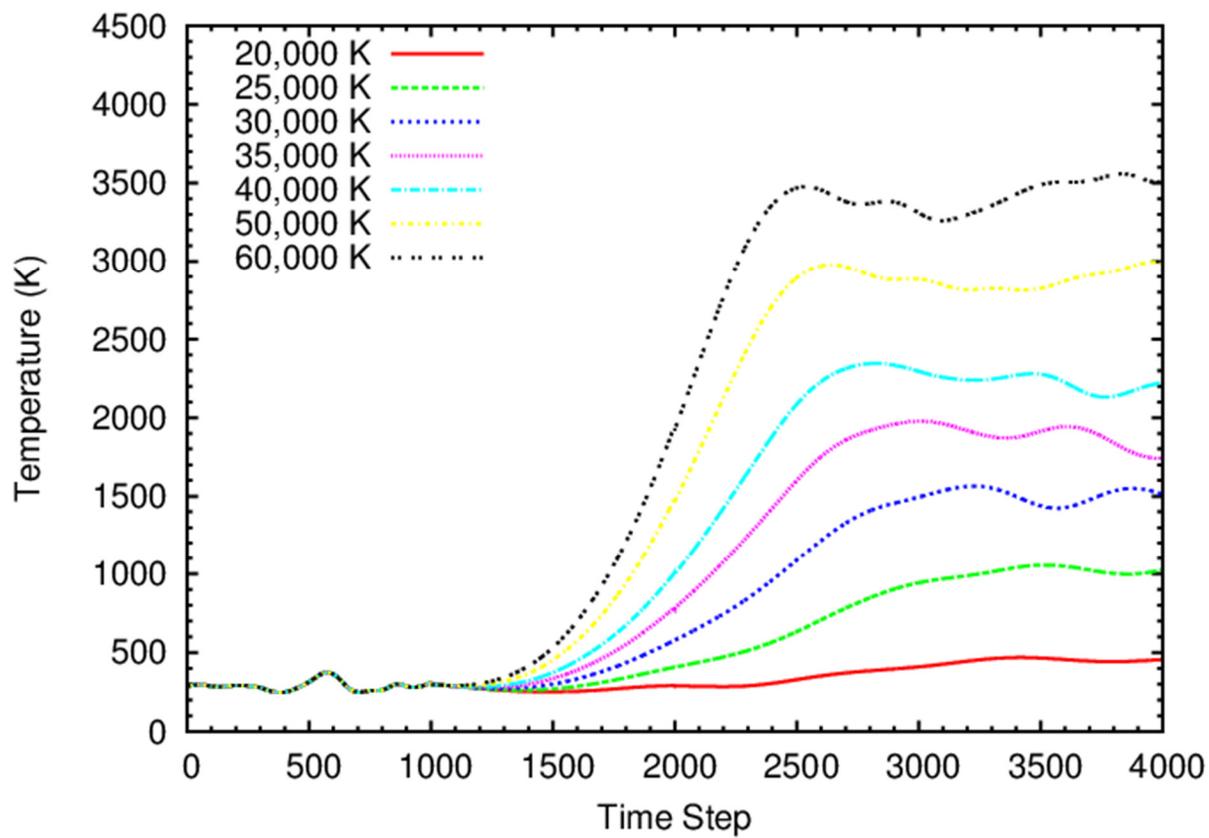

**Fig. 1**

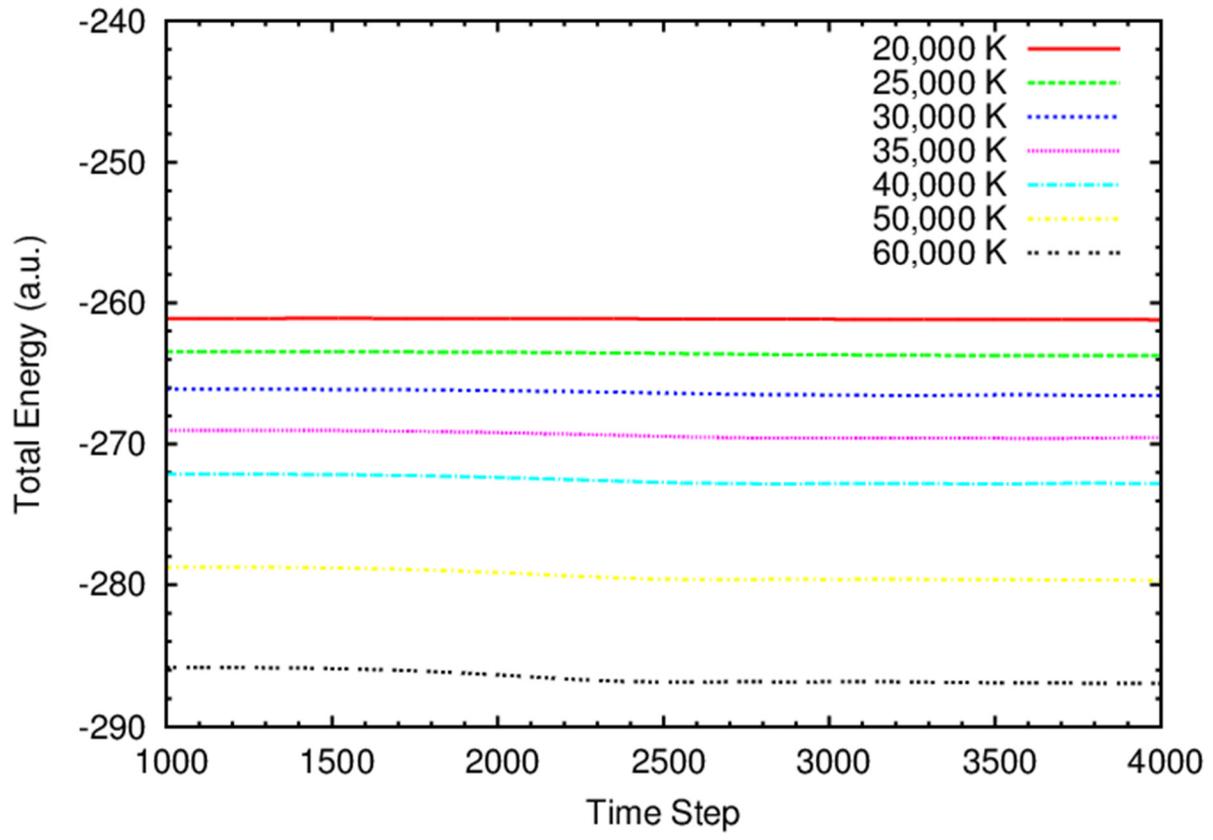

**Fig. 2**

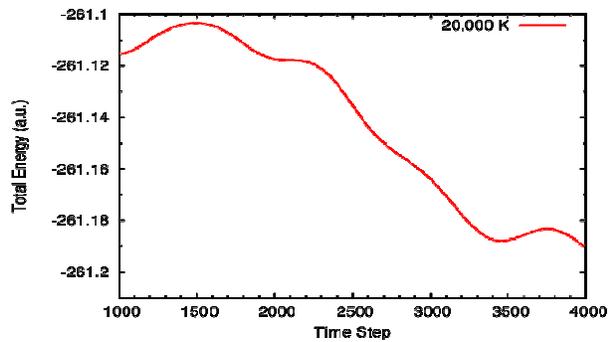
(a) 20,000 K

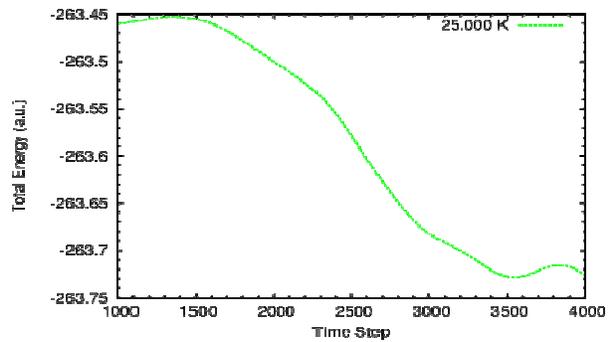
(b) 30,000 K

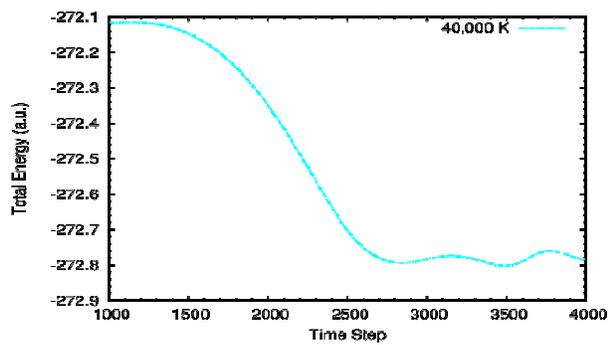
(c) 40,000 K

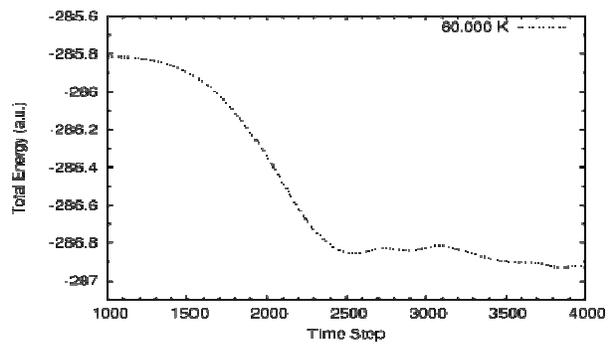
(d) 60,000 K

**Fig. 3**

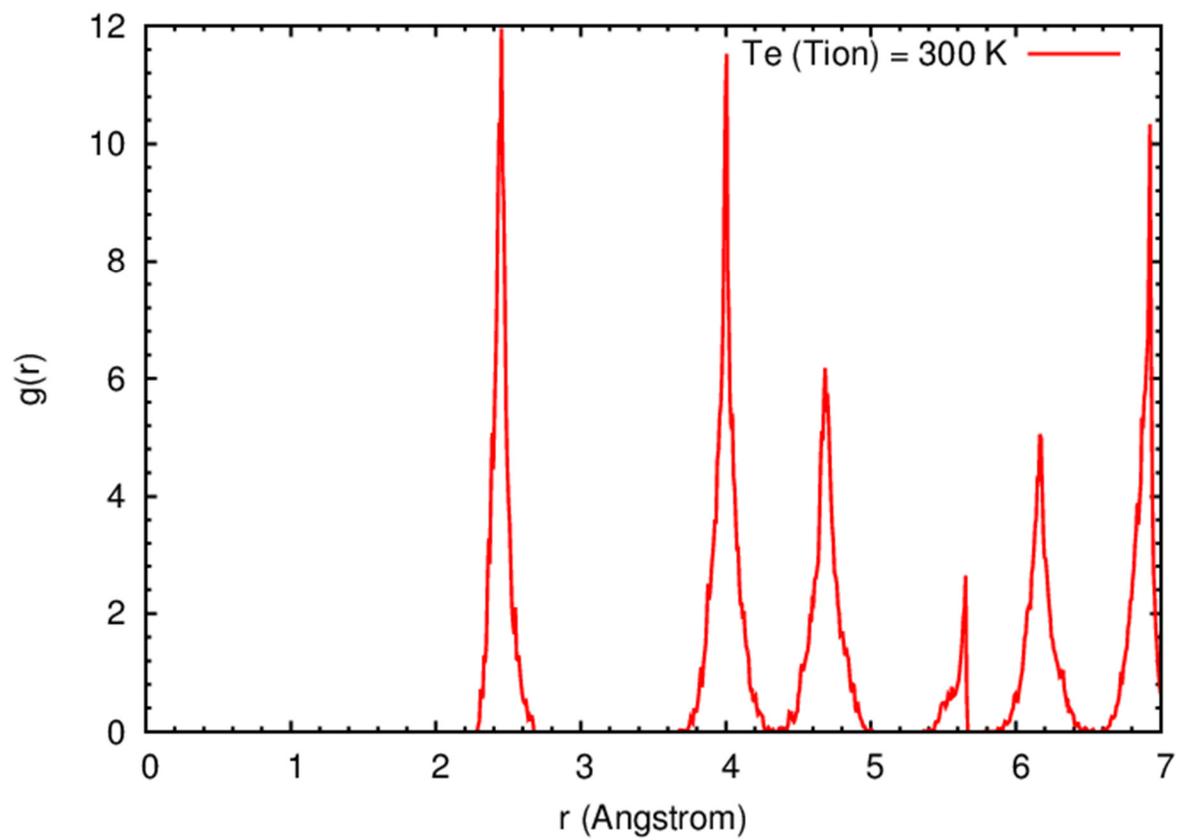

**Fig. 4**

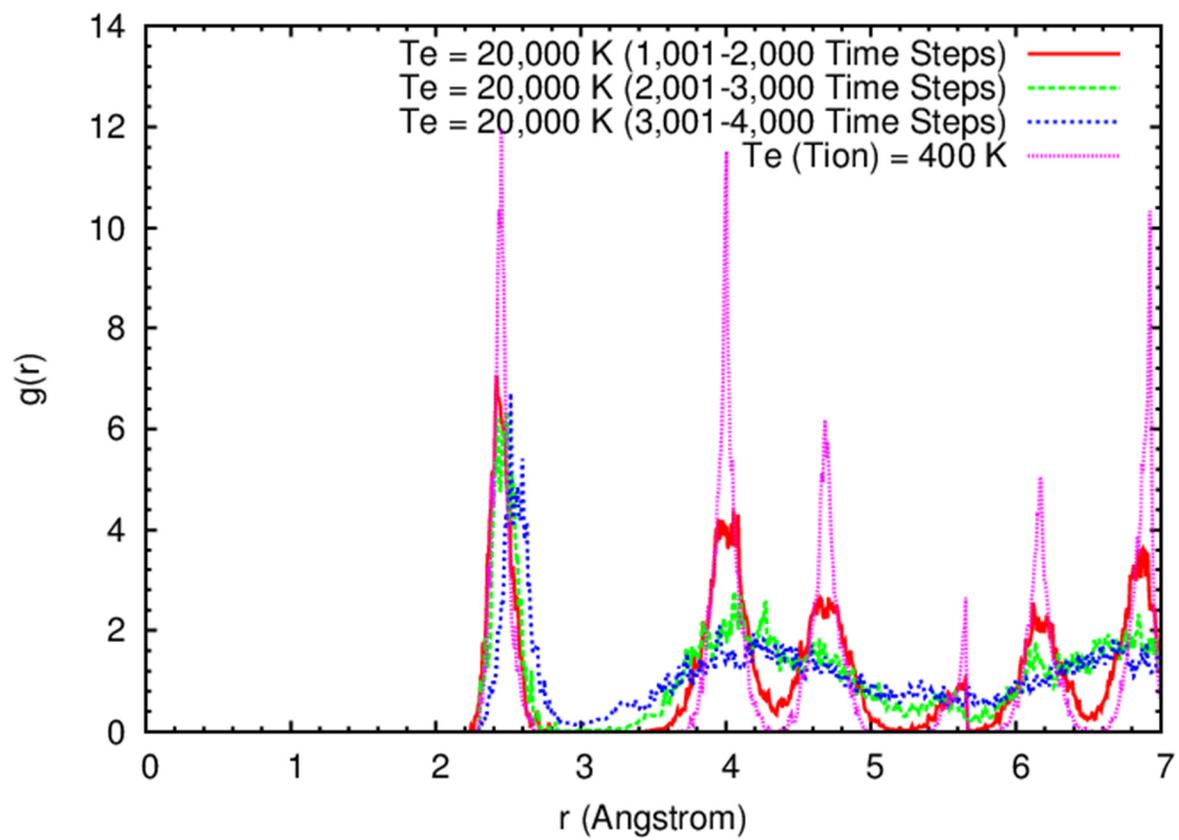

**Fig. 5**

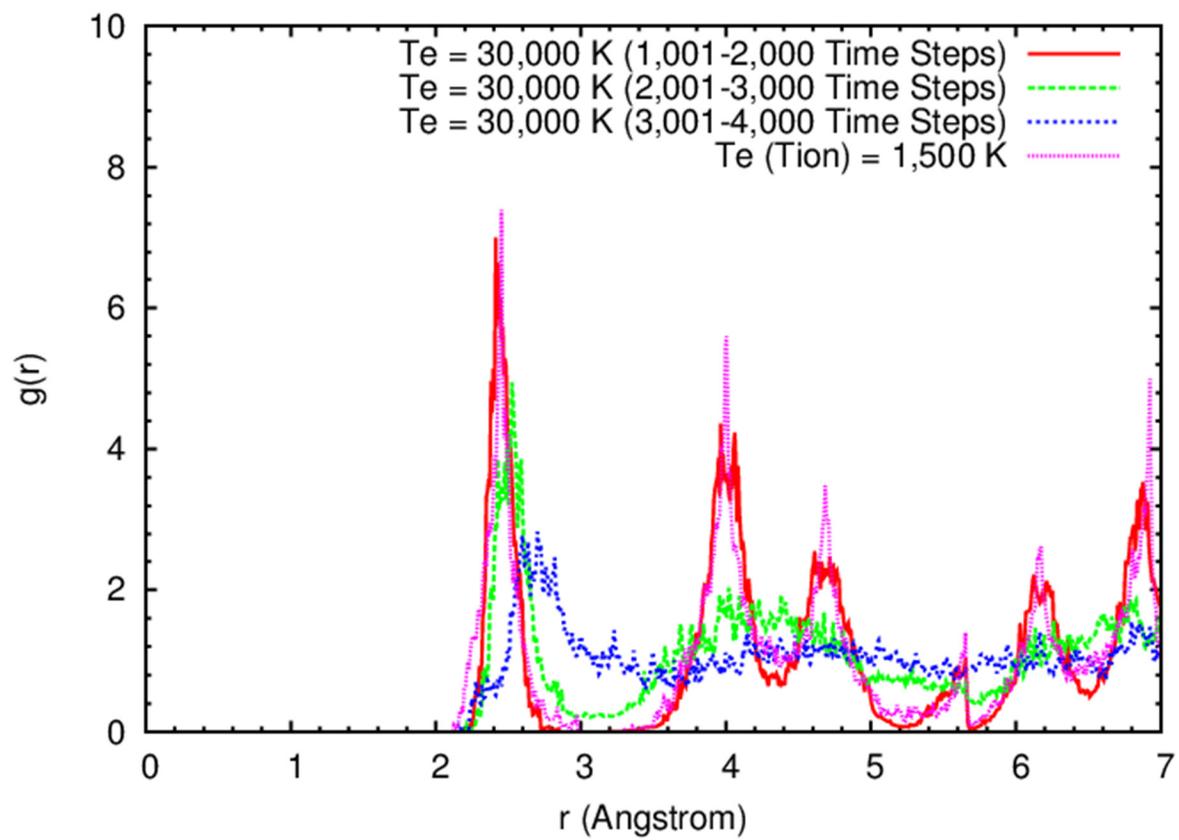

Fig. 6

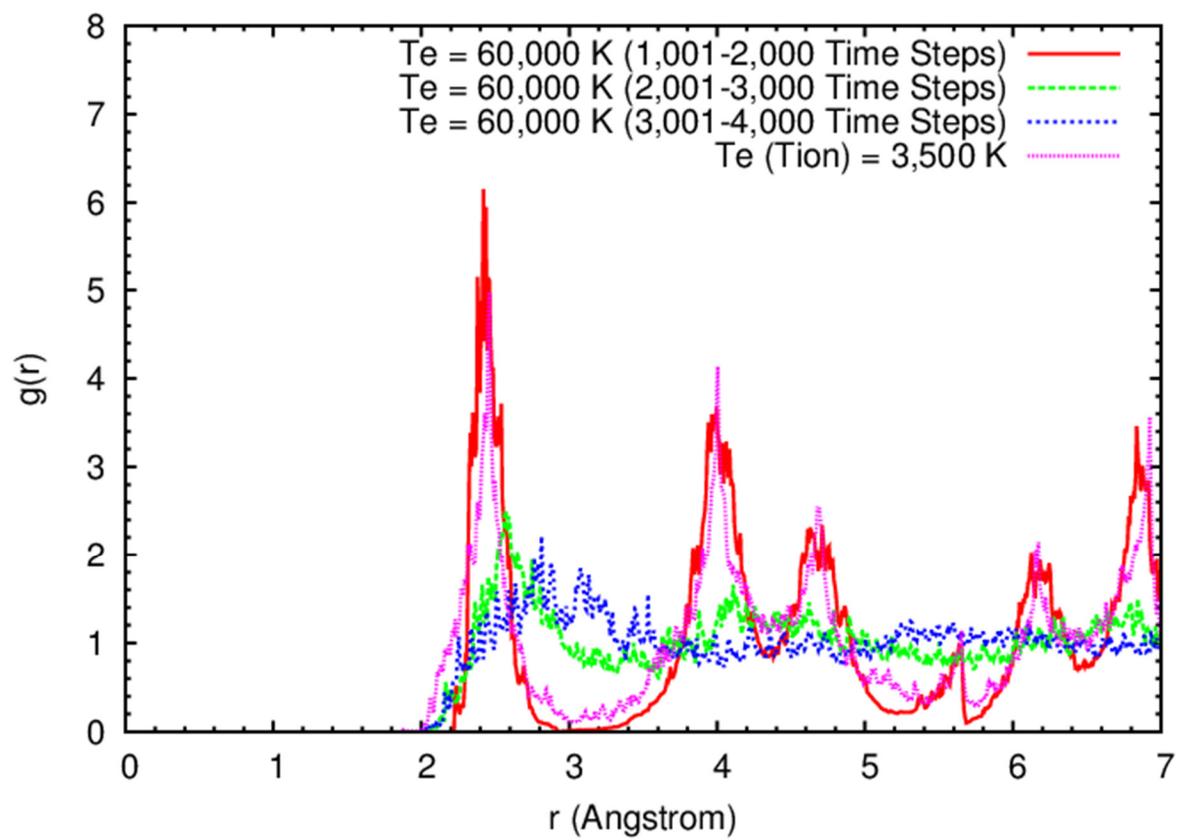

**Fig. 7**

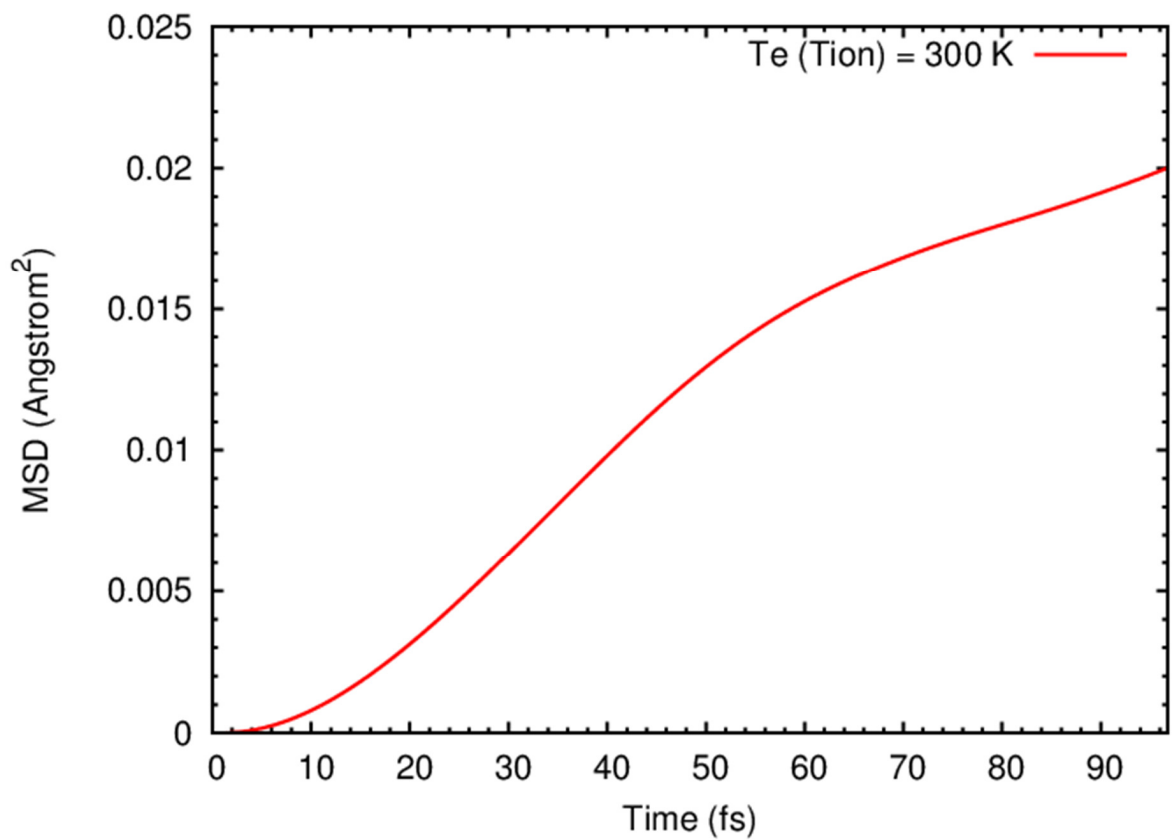

**Fig. 8**

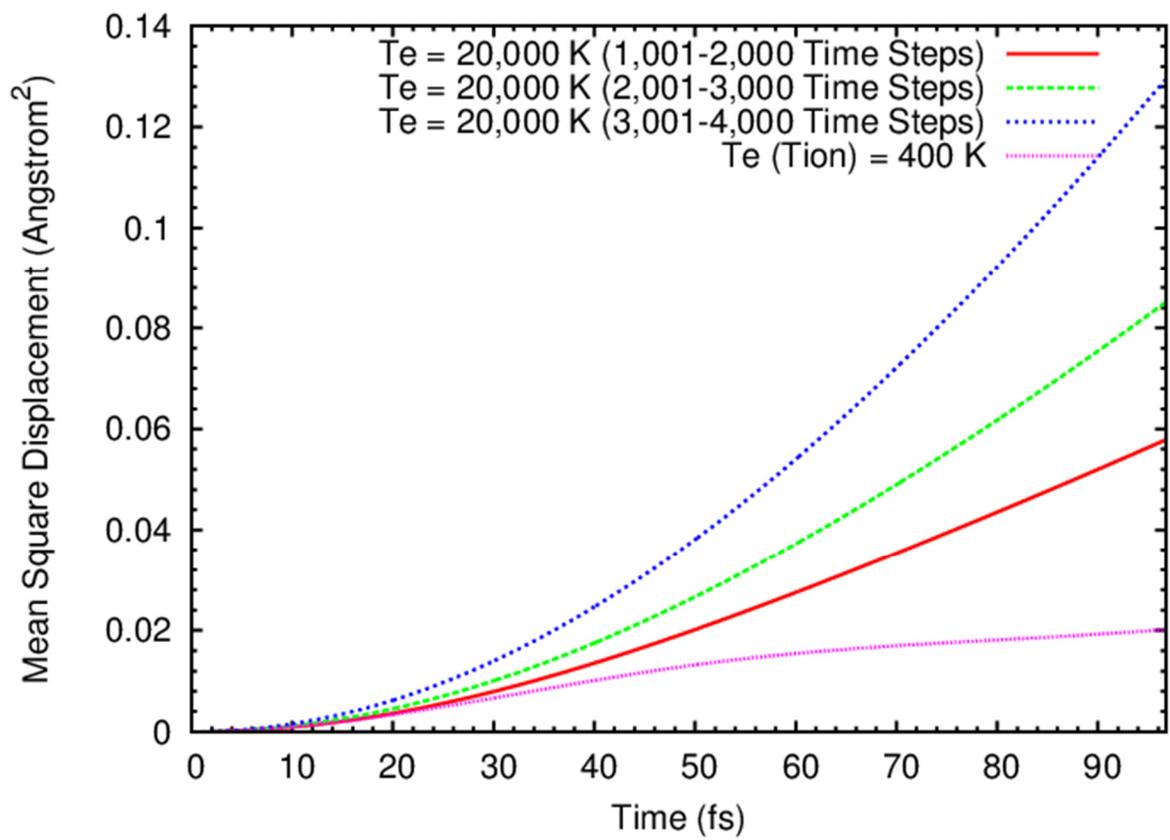

**Fig. 9**

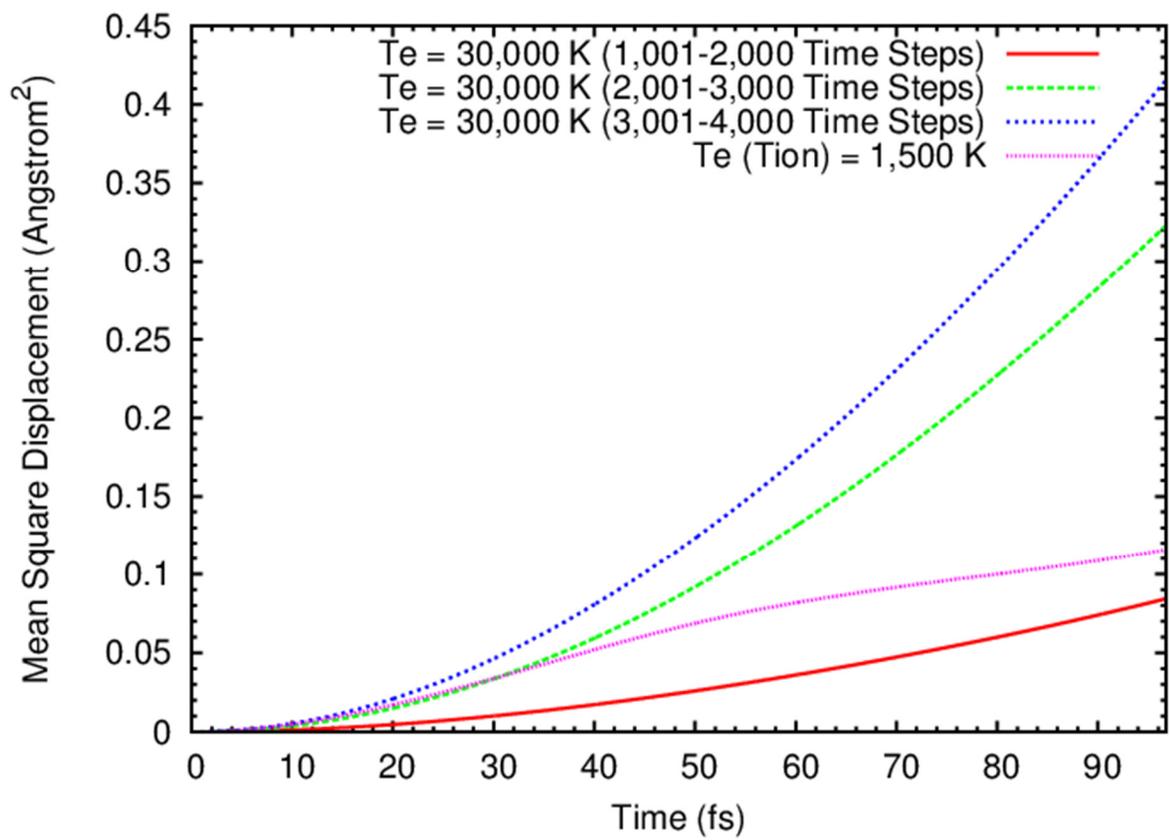

**Fig. 10**

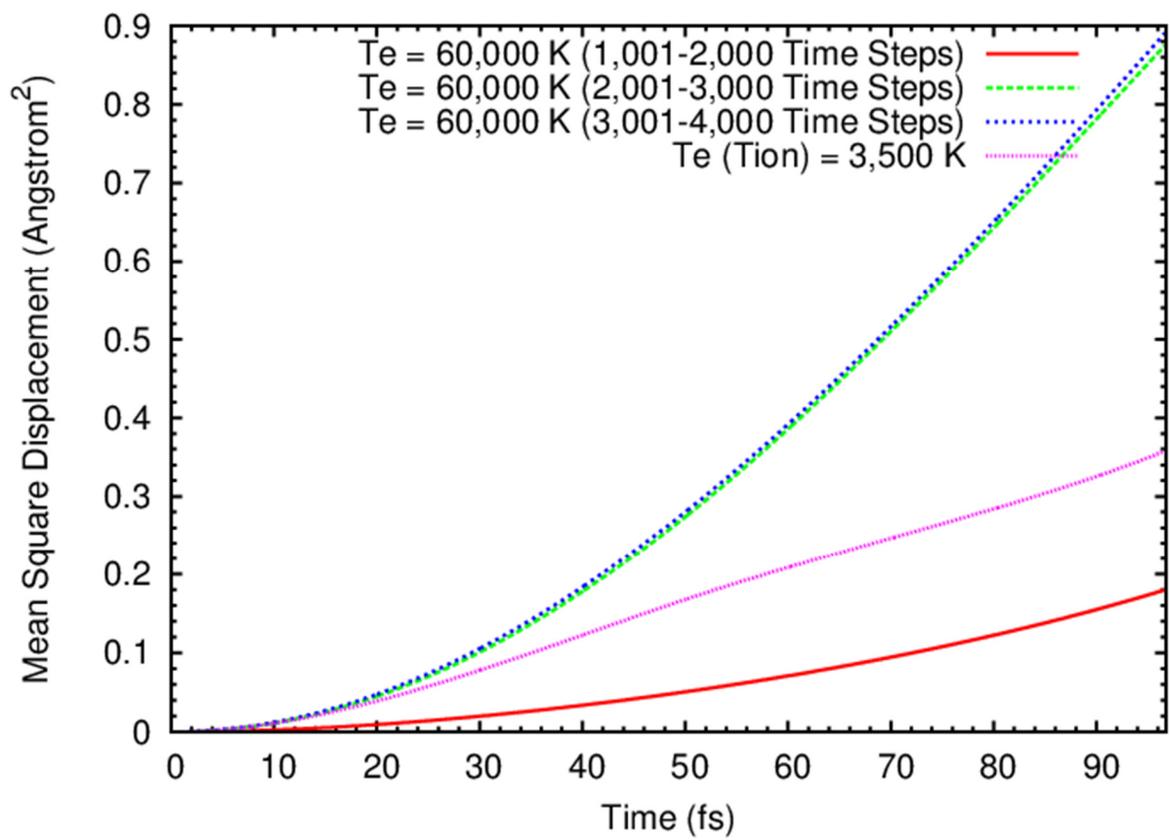

**Fig. 11**

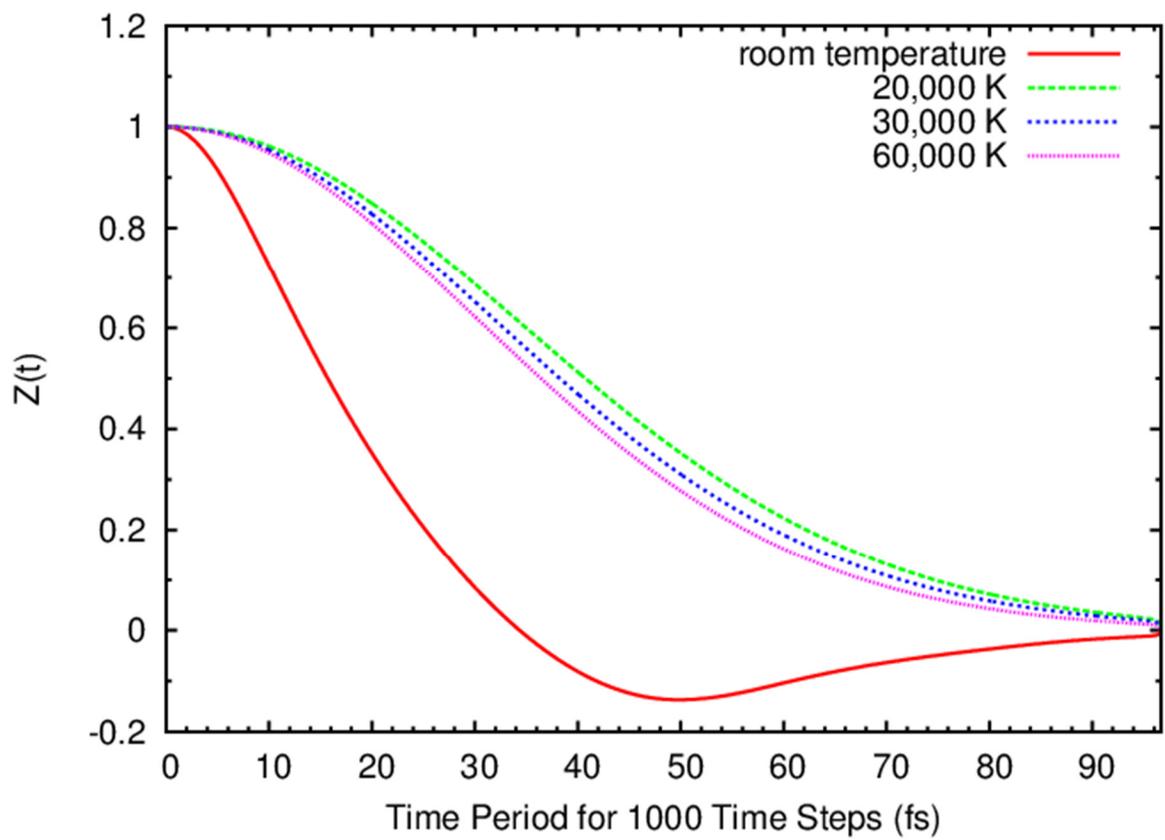

**Fig. 12**

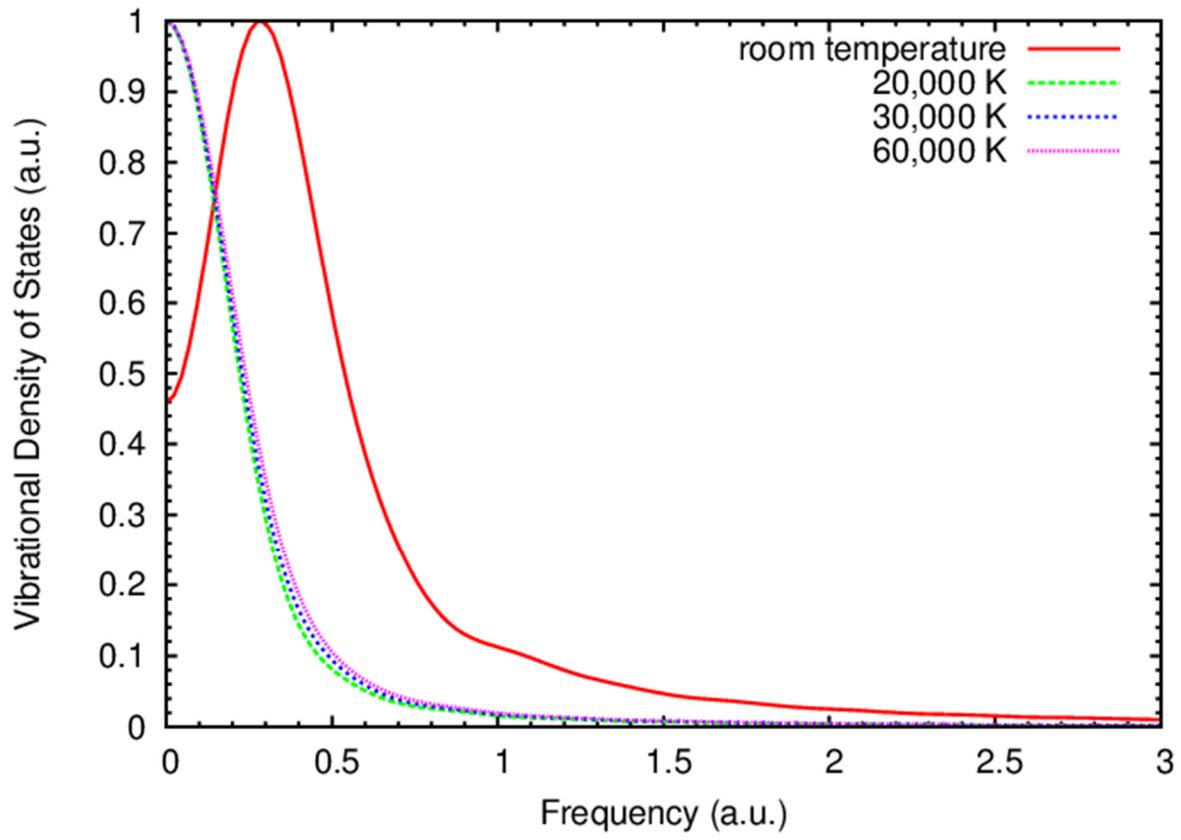

**Fig. 13**